\documentclass[a4paper,11pt,dvips]{article}
\usepackage{graphicx}
\usepackage{amsmath}
\usepackage{amssymb}
\usepackage{latexsym}
\usepackage[colorlinks]{hyperref}
\usepackage{color}
\usepackage{float}
\usepackage{cite}
\usepackage{makeidx}
\usepackage{colortbl}
\newcommand{\bra}{\begin{array}}
\newcommand{\era}{\end{array}}
\newcommand{\beq}{\begin{equation}}
\newcommand{\eeq}{\end{equation}}
\newcommand{\beqar}{\begin{eqnarray}}
\newcommand{\eeqar}{\end{eqnarray}}

\def\BC{\bb C}
\def\_\BC{\bbi C}

\def\( {\left(}
   \def\) {\right)}
\def\[ {\left[}
\def\] {\right]}
\def\no2 {{\textstyle{n\over 2}}}

\def\ra {{\rangle}}

\newcommand{\si}{\sigma}

\newcommand{\lga}{\longrightarrow}

\newcommand{\lb}{\label}

\begin{document}
\begin{titlepage}
\setcounter{page}{1}
\renewcommand{\thefootnote}{\fnsymbol{footnote}}

\begin{flushright}
ucd-tpg:1103.07\\
%arXiv:yymm.xxxx
\end{flushright}

\vspace{5mm}
\begin{center}

{\Large \bf {Tunneling of Graphene Massive Dirac Fermions \\ through
a Double Barrier}} %\\ and Zeeman Effect}} % for Graphene Systems}}

\vspace{5mm}
{\bf Hocine Bahlouli}$^{a,b}$, {\bf El Bou\^azzaoui Choubabi}$^{c}$,
{\bf Ahmed Jellal$^{a,c,d}$\footnote{\sf ajellal@ictp.it -- a.jellal@ucd.ac.ma}}
and {\bf Miloud Mekkaoui}$^{c}$

\vspace{5mm}

{$^a$\em Saudi Center for Theoretical Physics, Dhahran, Saudi Arabia}

{$^b$\em Physics Department,  King Fahd University
of Petroleum $\&$ Minerals,\\
Dhahran 31261, Saudi Arabia}

{$^{c}$\em Theoretical Physics Group,  %Department of Physics,
Faculty of Sciences, Choua\"ib Doukkali University},\\
{\em PO Box 20, 24000 El Jadida,
Morocco}

{$^d$\em Physics Department, College of Science, King Faisal University,\\
PO Box 380, Alahsa 31982, Saudi Arabia}

\vspace{3cm}

\begin{abstract}

We study the tunneling  of Dirac fermions in graphene through a double barrier
potential. This is  allowing the carriers to have an effective mass inside the barrier as
generated by a lattice miss-match with the boron nitride substrate.
The consequences of this gap opening on the transmission are investigated and
the realization of resonant tunneling conditions is  analyzed.

\vspace{3cm}



\end{abstract}
\end{center}
\end{titlepage}

\newpage
%%%%%%%%%%%%%%%%%%%%%%%%%%%%%%%%%%%%%%%%%%%%%%%%%%%
\section{Introduction }
%%%%%%%%%%%%%%%%%%%%%%%%%%%%%%%%%%%%%%%%%%%%%%%

After the experimental realization of graphene in 2005, this system became an attractive
subject not only to experimentalists but also to theoretical physicists. This is partially
due to the relativistic-like behavior of its massless carriers. In addition to its
anomalous quantum Hall effect~\cite{novoselov,zhang}, graphene provides an example of condensed matter
systems where quantum electrodynamics tools can be applied~\cite{katsnelson}. These new developments offered
a condensed matter laboratory to perform many investigations, which were solely available to high energy physicist
in the past. Because of this similarity with relativistic fermions, graphene systems triggered
the interest of the scientific community for studying massless Dirac fermions in two-dimensional spaces.

One of the characteristics of Dirac fermions in graphene is their ability to tunnel through very
high potential barriers with unit probability~\cite{NaturePhys,Bai07}, contrary to the usual intuition.
This so-called Klein tunneling of chiral particles has long ago been proposed in the framework of quantum
electrodynamics~\cite{Klein29,Calogeracos,Zuber} and was recently
observed experimentally \cite{stander, young}. However, as appealing as the Klein tunneling may sound from the fundamental research point
of view, its presence in graphene is unwanted when it comes to applications of graphene because space confinements
of the carriers is of great importance in nanoelectronics. One way to overcome these difficulties is through generating
a gap in the energy spectrum or equivalently to generate a mass term in the associated Dirac equation.

Experimentally it has been shown that one can generate a gap in the graphene spectrum if a graphene sheet is
deposited on top of hexagonal boron nitride (BN)~\cite{Giovannetti07}. In graphene the carbon-carbon distance is 1.42 \AA,
when deposited on a substrate of boron nitride, a band gap insulator with a boron to nitrogen distance of the order
of 1.45 \AA~\cite{Zupan71} creates a lattice miss-match which is at the origin of gap opening in the graphene sheet.
It was shown that the most stable configuration obtains when one carbon is on top of a boron and the other carbon
in the unit cell is centered above a BN ring, the value of the induced gap is of the order of 53 m{\ttfamily eV}.
Depositing graphene on a metal surface with a BN buffer layer leads  to $n$-doped graphene with an energy gap of 0.5
{\ttfamily eV}~\cite{Lu07}.

Theoretically, tunneling phenomena through graphene deposited on SiO$_2$-BN substrate in zero magnetic field was discussed~\cite{peres}.
In that work it was assumed that it was possible to manufacture a thin layer with SiO$_2$-BN interfaces, on top of which a graphene flake
was deposited. This arrangement induces spatial regions where graphene has a vanishing gap intercalated with regions where BN will cause
a finite gap. The graphene physics was considered in two different regions: the $k-$region, where the graphene sheet is standing on top
of SiO$_2$, and a $q-$region where the sheet stands over a BN sheet hence causing a mass-like term and inducing an energy gap of value $2t'$. %(for all numerical purposes we use $t'=$0.1 {\ttfamily eV}).
The effect of chiral electrons in graphene through these region was studied, the electronic spectrum changes
from the usual linear dispersion to a hyperbolic dispersion, due to the presence of a gap. It was shown that contrary
to the tunneling through a potential barrier, the transmission of electrons was, in this case, smaller than the one for
normal incidence. These
results were extended to the case of a magnetic field case \cite{zitt} and different generalizations were obtained.
%they claimed that
%their mechanism may be useful for designing electronic
%devices made of graphene.

Motivated by the above reasons and in particular the investigation made in~\cite{peres} and \cite{zitt},
we would like to extend the study of tunneling phenomena through a graphene sheet deposited on
SiO$_2$-BN layer in the presence of an external magnetic field $B$ and
double barrier potentials.
%We will investigate how
%the conclusions reached in the case $B=0$ and in the absence of barrier in~\cite{peres} are affected by
%the presence of a magnetic field.
Note that, actually there were many theoretical investigations on tunneling of electrons in graphene in the presence of magnetic field
%Note that, there are actually some works on the tunneling effect in terms of
%magnetic fields
e.g.~\cite{demartino,shytov,peeters,mr}. However, they did not consider a graphene system deposited on SiO$_2$-BN and hence their graphene system was gapless.

More precisely, we consider a system made of two different regions, where the second is characterized by an energy gap $t'$,
in a perpendicular  magnetic field. The energy spectra are obtained in both regions in terms of two different Landau levels and $t'$.
%and determine the reflexion and transmission
%coefficients in terms of $B$. For this, we start by evaluating the eigenvalue
%solution for each regions and determine the corresponding wavefunctions.
For concreteness, we consider a barrier in magnetic field and study the tunneling effect. Indeed, from
the continuity condition we obtain different solutions, which allowed us to explicitly determine
the reflection and transmission coefficients in different regions those are then shown to obey
the probability conservation law. We study the transmission behavior in terms of the energy ratio ${E\over t'}$ and
discuss the possibility of resonant tunneling.

{ We  emphasis that the inhomogeneous magnetic field $B$ discussed in our manuscript is applied externally. It can be created for instance by depositing a type-I superconducting film on top of the system and remove a strip $- d_1 < x < d_1$ of the superconductor and apply a perpendicular magnetic field. This patterning technique of creating the desired magnetic field profile was proposed  by Matulis {\it et al.} \cite{matulis}.
%by A. Matulis, F.M. Peeters, and P. Vasilopoulos: "Wavevector-dependent tunneling through magnetic barriers", Phys. Rev. Lett. 72, 1518-1521 (1994).
One of the interesting features of such inhomogeneous magnetic field profile is that it can bind electrons, contrary to the usual potential step. Such a step magnetic field will indeed result in electron states that are bound to the step $B$-field and that move in one direction along the step. Thus there is a current along the $y$-direction but that is a very small effect and it  is not relevant for our problem (those electrons have $k_x = 0$). Indeed, we consider free electron states that have in general $k_x$ non zero, because otherwise they will not tunnel.
A recent work studied double barriers with magnetic field for graphene, i.e. no mass term, can be found in
 \cite{masir}.}
 %in the following reference. M. Ramezani Masir, P. Vasilopoulos, and F.M. Peeters: "Fabry-Perot resonances in graphene microstructures: influence of a magnetic field", Phys. Rev. B 82, 115417 (2010).}

The present paper is organized as follows. In section 2, we formulate our problem
to include different part in the Hamiltonian describing the system made of graphene.
The corresponding solution of the energy spectrum will be the subject of section 3.
The  scattering problem for Dirac
fermions will be solved in section 4 where the transfer matrix will be explicitly
given. In section 5, we determine the reflection and transmission coefficients
using the associated current density. We conclude our work in
the final section.

%%%%%%%%%%%%%%%%%%%%%%%%%%%%%%%%%%%%%%%%%%%%%%%%%%%
\section{Problem formulation }
%%%%%%%%%%%%%%%%%%%%%%%%%%%%%%%%%%%%%%%%%%%%%%%

To deal with our task
let us first describe the geometry of our system, which is made of
five regions denoted by $j=1,2, \cdots 5$. Each region is characterized by its potential
and interaction with external sources. The potential profile is summarized in Figure 1:

 \begin{center}
 \includegraphics[width=4in]{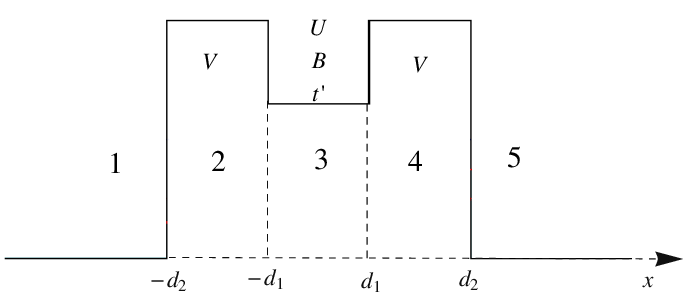}
\end{center}
%\begin{text}
{\sf {Figure 1: The energy diagram of the potential $V
(x)$ describing our double barrier along with its physical parameters.}}\\
%\end{text}
%\end{figure}
%%%%%%%%%%%%%%%

\noindent The double barrier potential $V(x)$ is defined by
%Mathematically, we have the following configuration for
%the potential $V(x)$
%The series of potential is described by the following equation
\begin{equation}
V(x)=
\left\{%
\begin{array}{ll}
    V, & \qquad \hbox{$d_{1}<|x|<d_{2}$} \\
     U, & \qquad \hbox{$|x|<d_{1}$} \\
    0, & \qquad \hbox{otherwise}. \\
\end{array}%
\right.
\end{equation}
According the region labels, we denote
%where $j=1,2, \cdots 5$ labels different potential regions and their
 the corresponding constant potential values  by $V_j$. We introduce a uniform
perpendicular magnetic field, along the $z$-direction, constrained to the well region between the two barriers and is defined by %$\vec{B}=B_{0}\vec{e_{z}}$
\begin{equation}
B(x,y)=B_{0}\Theta\left(d_{1}^{2}-x^{2}\right)
\end{equation}
where $B_{0}$ is the strength of the magnetic field within the strip located in the region
$|x|\leq d_{1}$ and $B=0$ otherwise, $\Theta$ is the Heaviside step function. Choosing the Landau
gauge and imposing %the potential vector $\overrightarrow{A} = (0,
  %Ay,  0)^{T}$ created by the magnetic field, The
continuity of the vector potential at the boundary to avoid nonphysical effects, we end up with the following vector potential
\begin{equation}
\qquad A_{y}(x)=\frac{c}{el_{B}^{2}}\times\left\{%
\begin{array}{ll}
    -d_{1}, & \qquad \hbox{$x<-d_{1}$} \\
    x, & \qquad \hbox{$ |x| <d_{1}$} \\
    d_{1}, & \qquad \hbox{$x>d_{1}$} \\
\end{array}%
\right.
\end{equation}
with the magnetic length defined by  $l_{B}=\sqrt{1/ B_{0}}$ in  the unit system $(\hbar=c=e=1)$.
%, $c$ being the speed of light and $e$ the electronic charge. $l_{B}=\sqrt{c/e B_{0}}$

Once we have defined the potential parameters relevant to all regions, we introduce the Hamiltonian for one-pseudospin component describing
our system.  In region $j=1,2,4,5$ it can be written as
\begin{equation}\label{5}
H_{j}=v_{\sf F} \vec{\sigma} \cdot \vec{\pi}+ V_{j} %{\mathbb I}_{2}
\end{equation}
while in region $(3)$ we have
\begin{equation}\label{31}
H_{3}=v_{\sf F} \vec{\sigma} \cdot \vec{\pi}+
U %{\mathbb I}_{2}
+t^{'}\sigma_{z}
\end{equation}
where $v_{\sf F}$ is the Fermi velocity,
$\vec{\pi}=\vec{p}+e\vec{A}/c$
is the two-component momentum, %with the canonical momentum $\vec{p}$,
$t'$ is the mass term,
 $ \vec{\sigma}= (\sigma_{x}, \sigma_{y})$ and $\si_z$
are the usual Pauli matrices.
To proceed further, we need to find the solutions of the corresponding Dirac equation and
therefore the energy spectrum.
%which will be the subject of the next section.

%%%%%%%%%%%%%%%%%%%%%%%%%%%%%%%%%%%%%%%%%%%%%%%%%%%%%%%%%
\section{Energy spectrum}
%%%%%%%%%%%%%%%%%%%%%%%%%%%%%%%%%%%%%%%%%%%%%%%%%%%%%%%%%

We determine the eigenvalues and eigenspinors corresponding
to the Hamiltonian $H_{j}$ in different regions. Indeed, %the Dirac Hamiltonian describing
from (\ref{5}) we can write in regions (1, 2, 4, 5) the Hamiltonian as
\begin{equation}
H_{j}= v_{\sf F} \left(%
\begin{array}{cc}
  V_{j}/ v_{\sf F} & p_{jx}-ip_{y} + i\frac{d}{l_{B}^{2}} \\
p_{jx}+ip_{y}-i\frac{d}{l_{B}^{2}} & V_{j}/v_{\sf F} \\
\end{array}%
\right)
\end{equation}
where the parameter $d$ is defined by
\begin{equation}
d=\left\{%
\begin{array}{ll}
    d_{1}, &\qquad  \hbox{$x<-d_{1}$} \\
    -d_{1}, & \qquad \hbox{$x>d_{1}$}. \\
\end{array}%
\right.
\end{equation}
To solve the eigenvalue problem,
we can separate variables and write the eigenspinors as
%The time independent Dirac equation for the spinor
\beq
\psi_{j}(x, y)= \varphi_j (x) \chi(y)
\eeq
 where $\varphi_j (x)=(\varphi_{j+} \ \ \varphi_{j-})^{t}$, $\varphi_{j\pm}$ are the upper and lower spinor components,
and $\chi(y)= e^{ik_yy}$, with $k_y$ a real parameter that stands for the wave number of the excitations along
the $y$-axis. After rescaling  energy
$\epsilon = E/v_{\sf F}$ and potential $v_j = V_{j}/v_{\sf F}$, the resulting reduced time independent Dirac equation is given by
\begin{equation}
\left(%
\begin{array}{cc}
  0 & p_{jx}-ip_{y} +i \frac{d}{l_{B}^{2}} \\
  p_{jx}+i p_{y}-i\frac{d}{l_{B}^{2}} & 0 \\
\end{array}%
\right)\left(%
\begin{array}{c}
  \varphi_{j+} \\
  \varphi_{j-} \\
\end{array}%
\right)=(\epsilon-v_{j})\left(%
\begin{array}{c}
  \varphi_{j+} \\
  \varphi_{j-} \\
\end{array}%
\right).
\end{equation}
This eigenproblem can be written as two linear differential equations of the from
\begin{eqnarray}
%\left\{%
%\begin{array}{ll}
   && \left(p_{jx}-ip_{y} +i \frac{d}{l_{B}^{2}}\right)\varphi_{j-}=(\epsilon-v_{j})\varphi_{j+}\\ %, %& \hbox{} \\
    && \left(p_{jx}+ip_{y}- i\frac{d}{l_{B}^{2}}\right)\varphi_{j+}=(\epsilon-v_{j})\varphi_{j-}
\end{eqnarray}
where $p_{jx}$ is the $x$-component of the momentum operator in the $j$-th potential region.
The above equation can be solved, using the unit system $(\hbar=c=e=1)$, to get the  energy eigenvalues
\begin{equation}
\epsilon-v_{j}=s_{j}
\sqrt{(k_{jx})^{2}+\left(k_{y}-\frac{d}{l_{B}^{2}}\right)^{2}}
\end{equation}
where $s_{j}=\mbox{sign}(\epsilon-v_{j})$ is the usual sign function, equal to $\pm 1$ for
a positive and negative argument, respectively. It is clear that  the wave vector along the
$x$-direction can be written as
\begin{equation}
k_{jx}=\sqrt{(\epsilon -v_{j})^{2}-\left(k_{y}-\frac{d}{ l_{B}^2}\right)^{2}}.
\end{equation}

Consider now a particle-like scattering state $(\epsilon >0)$, the particle is incident from the left side in
the $j$-th potential region, with incoming wave vector $k_{j}=(k_{jx}, k_{y})$ and position $r=(x, y)$.
Thus, the incoming spinor in the $j$-th potential region is given by
\begin{equation}
\psi_{j}(x,y)=\frac{1}{\sqrt{2}}\left(
\begin{array}{c}
1 \\
 z_{j}\end{array}\right)e^{i k_{j}.r}
\end{equation}
where $z_j$ is given by
\begin{equation}
%z_{p_{xi}}=
z_{j}=s_{j}\ \frac{k_{jx}
+i\left(k_{y}-\frac{d}{l_{B}^2}\right)} {\sqrt{(k_{jx})^{2}
+\left(k_{y}-\frac{d}{l_{B}^2}\right)^{2}}}=s_{j}
e^{i\phi_{j}}
\end{equation}
and  the phase is $\phi_{j}=\arctan\left(\frac{k_{y}-\frac{d}{l_{B}^{2}}}{k_{jx}}\right)$. The sign
$s_{j}= \pm 1$ again refers to conduction and valence bands in
region $j= 1,2,4,5$. The corresponding left and right moving spinors are defined by
\begin{eqnarray}
\psi_{j+}(x,y) &=& \frac{1}{\sqrt{2}}\left(
\begin{array}{c}
1 \\
 z_{j}\end{array}\right)e^{i(k_{jx} x +k_{y} y)}\lb{15}\\
\psi_{j-}(x,y) &=& \frac{1}{\sqrt{2}}\left(
\begin{array}{c}
1 \\
 -z_{j}^*\end{array}\right)e^{i (-k_{jx} x +k_{y} y)}.\lb{16}
\end{eqnarray}

To complete the solutions of the energy spectrum, we consider now region 3
%the Hamiltonian describing region 3
where $|x|\leq d_{1}$, %can be written as
%\begin{eqnarray}
%H=\upsilon_{F}\overrightarrow{\sigma}. \overrightarrow{\pi}+V_{2}
%I_{2}+t^{'}\sigma_{z}
%\end{eqnarray}
$\pi_{x}=p_{x}$, $\pi_{y}=p_{y}+A_{y}$, and
vector potential is
 $A(x)=x/l_{B}^{2}$,
then (\ref{31}) becomes
\begin{equation}
H_3=\left(%
\begin{array}{cc}
  U/v_{\sf F} +t^{'}/v_{\sf F} & -i\partial_{x} -i k_{y} -i \frac{x}{l_{B}^{2}}\\
 -i\partial_{x}+ ik_{y} +i\frac{x}{l_{B}^{2}} &  U/v_{\sf F}-t^{'}/v_{\sf F}\\
\end{array}%
\right).
\end{equation}
To diagonalize this Hamiltonian we proceed by defining the usual bosonic operators
\begin{eqnarray}
a=\frac{l_{B}}{\sqrt{2}} \left(\partial_{x}+k_{y}+\frac{x}{l_{B}^{2}}\right),
\qquad
a^{\dagger}=\frac{l_{B}}{\sqrt{2}} \left(-\partial_{x}+k_{y}+\frac{x}{l_{B}^{2}}\right)
\end{eqnarray}
which satisfy the commutation relation  $[a, a^{\dagger}]=1$. Rescaling our energies
$\mu= t^{'}/v_{\sf F}$ and $u=U/v_{\sf F}$, we can write the Hamiltonian in terms of $a$ and $ a^{\dagger}$ as follows
\begin{equation}
H_3=v_{\sf F}
\left(%
\begin{array}{cc}
  u+\mu & -i\frac{\sqrt{2}}{l_{B}}a \\
  i\frac{\sqrt{2}}{l_{B}}a^{\dagger}  &  u-\mu \\
\end{array}%
\right).
\end{equation}
The eigenvalue equation for the spinor $ \psi_{3}(x, y)$, that is
$H_3 \psi_{3}(x, y)=E\psi_{3}(x, y)$
gives
$\psi_{3}(x, y)=\varphi_3(x) \chi(y)$
with $\chi(y)=e^{ik_{y}y}$ and $\varphi_3(x)=
(\varphi_{3+}\ \ \varphi_{3-})^{t}$ verifies
\begin{equation}
 \left(%
\begin{array}{cc}
  u+\mu & -i\frac{\sqrt{2}}{l_{B}}_{c}a \\
  +i\frac{\sqrt{2}}{l_{B}}_{c}a^{\dagger}  &  u-\mu \\
\end{array}%
\right)\left(%
\begin{array}{c}
  \varphi_{3+} \\
  \varphi_{3-} \\
\end{array}%
\right)=\epsilon\left(%
\begin{array}{c}
  \varphi_{3+} \\
  \varphi_{3-} \\
\end{array}%
\right)
\end{equation}
where  again $\epsilon= E/v_{\sf F}$. The eigenproblem can then be rewritten as two relations between the two spinor
components
\begin{eqnarray}
  &&(u+\mu)\varphi_{3+}-i\frac{\sqrt{2}}{l_{B}}a\varphi_{3-}=\epsilon\varphi_{3+}\\
  && i\frac{\sqrt{2}}{l_{B}}a^{\dagger}\varphi_{3+}+
  (u-\mu)\varphi_{3-}=\epsilon\varphi_{3-}
\end{eqnarray}
which lead to an eigenvalue equation for the operator $a a^{\dagger}$
and
 $\varphi_{3+}$
\begin{equation}
\left[(\epsilon-u)^{2}-\mu^{2}\right]\varphi_{3+}=\frac{2}{l_{B}^{2}}a
a^{\dagger}\varphi_{3+}.
\end{equation}
It is clear that $\varphi_{3+}$ is an eigenstate of the number
operator %$\widehat{N}
$N=a^{\dagger}a$ and therefore we identify
$\varphi_{3+}$ with the eigenstates of the harmonic oscillator
$|n-1\rangle$, namely
\begin{equation}
 \varphi_{3+} \sim \mid n-1\rangle
\end{equation}
and the corresponding energy spectrum reads
\begin{equation}
(\epsilon-u)=\pm \epsilon_{n}=\pm\frac{1}{l_{B}}\sqrt{(\mu
l_{B})^{2}+2n}.
\end{equation}
The second spinor component can now be written as follows
\begin{equation}
\varphi_{3-}=\frac{i\sqrt{2}a^{\dagger}}{(\epsilon-u)l_{B}+\mu
l_{B}}\mid n-1\ra=
       \frac{i\sqrt{2n}}{(\epsilon-u)l_{B}+\mu l_{B}} \mid n\ra
\end{equation}
where $ \sqrt{2n}=\sqrt{(\epsilon_{n}l_{B})^{2}-(\mu l_{B})^{2}}$,
which gives
\begin{equation}
\varphi_{3-}=\pm i\sqrt{\frac{\epsilon_{n}\mp \mu
}{\epsilon_{n}\pm \mu}} \mid n \ra.
\end{equation}
Where the $\pm$ signs correspond to positive and negative energy solutions, respectively.
After normalization we obtain the eigenspinors
for positive and negative energies
\begin{equation}
(\varphi)^{\pm}_{3n}=\frac{1}{\sqrt{2}}\left(%
\begin{array}{c}
  \sqrt{\frac{\epsilon_{n}\pm \mu}{\epsilon_{n}}} \mid n-1\ra \\
  \pm i\sqrt{\frac{\epsilon_{n}\mp \mu}{\epsilon_{n}}} \mid n\ra \\
\end{array}%
\right).
\end{equation}
By introducing the parabolic cylinder functions
$D_{n}(x)=2^{-{n}/{2}}e^{-\frac{x^{2}}{4}}H_{n}\left(\frac{x}{\sqrt{2}}\right)$,
the solution in the barrier region $\mid x\mid\leq d_{1}$ can be expressed as
\begin{equation}\lb{29}
\psi_{3n}(x, y)=\frac{1}{\sqrt{2}} %\sum_{\pm}c^{\pm}
\left(%
\begin{array}{c}
 \sqrt{\frac{\epsilon_{n}\pm \mu}{\epsilon_{n}}}
 D_{\left[(\epsilon_{n}l_{B})^{2}-(\mu l_{B})^{2}
 \right]/2-1}
 \left[\pm \sqrt{2} \left(\frac{x}{l_{B}}+k_{y}l_{B}\right)\right] \\
  \pm i\frac{\sqrt{2/l_B}}{\sqrt{\epsilon_{n}\left(\epsilon_{n}\pm \mu \right)}}
  D_{\left[(\epsilon_{n}l_{B})^{2}-(\mu l_{B})^{2}\right]/2}
  \left[\pm \sqrt{2}\left(\frac{x}{l_{B}}+k_{y}l_{B}\right)\right] \\
\end{array}%
\right)e^{ik_{y}y}.
\end{equation}
We should note that the above $ \pm $ signs stand for waves traveling to right and left, respectively,
along our usual convention for traveling waves $e^{\pm i k_{jx}x}$.
In this context the full solution in region 3 is given by a linear combination, $c_+$ times (\ref{29})
with the upper sign added to $c_-$ times (\ref{29}) with the lower sign where $c_{\pm}$
are arbitrary coefficients to be fixed by the boundqry conditions as exposed in the following section.
%with complex coefficients $c_{\pm}$.
The above results (\ref{15})-(\ref{16}) and (\ref{29}) summarizes the solutions of the energy spectrum in different
regions composing our graphene sheet. Next, we will show how to implement the
results obtained so far to study the tunneling of Dirac fermions through a double barrier potential.
%Having settled all ingreadient

%%%%%%%%%%%%%%%%%%%%%%%%%%%%%%%%%%%%%%%%%%%%%%%%%%%
\section{Boundary conditions}
%%%%%%%%%%%%%%%%%%%%%%%%%%%%%%%%%%%%%%%%%%%%%%%

It is straightforward to solve the scattering problem for Dirac
fermions. We assume that the incident wave propagates at an angle
$\phi_{1}$ with respect to the $x$-axis, where the superscript indicate the potential region. Let us first
write explicitly the eigenvalues and the corresponding eigenspinors
with reflected and transmitted parts in each region.
In the incident region for $x<-d_{2}$ (region 1), we have
\begin{eqnarray}
&& \epsilon = \sqrt{(k_{1x})^{2}+\left(k_{y}-\frac{d_{1}}{l_{B}^{2}}\right)^{2}}\\
%\end{equation}
%\begin{equation}
&& \psi_{1} = \frac{1}{\sqrt{2}}\left(
\begin{array}{c}
1 \\
 z_{1}\end{array}\right)e^{i(k_{1x} x +k_{y} y)}+r\frac{1}{\sqrt{2}}\left(
\begin{array}{c}
1 \\
 -z_{1}^*\end{array}\right)e^{i(-k_{1x} x +k_{y}
 y)}
\end{eqnarray}
where $r$ is the reflection amplitude. It is clear that  the shift in $k_{y}$ is due to our gauge
choice for the vector potential. It is  convenient to parameterize the momenta as follows
\begin{equation}
k_{1x}=\epsilon\cos\phi_{1},\qquad
k_{y}=\epsilon\sin\phi_{1}+\frac{d_{1}}{l_{B}^{2}}.
\end{equation}
In the barrier region $-d_{2} \leq x \leq-d_{1}$ (region 2), we have
\begin{eqnarray}
&&\epsilon-v =
s^{2}\sqrt{(k_{2x})^2+\left(k_{y}-\frac{d_{1}}{l_{B}^{2}}\right)^{2}}\\
%\end{equation}
%\begin{equation}
&&\psi_{2}  = \frac{a}{\sqrt{2}}\left(
\begin{array}{c}
1 \\
 z_{2}\end{array}\right)e^{i(k_{2x} x +k_{y} y)}+\frac{b}{\sqrt{2}}\left(
\begin{array}{c}
1 \\
 -z_{2}^*\end{array}\right)e^{i(-k_{2x} x +k_{y}
 y)}
\end{eqnarray}
where $a$ and $b$ are two wavefunction parameters and $v=V/v_F$. The corresponding, suitably parameterized, momentum is given by
\begin{equation}
k_{2x}=(\epsilon-v)\cos\phi_{2},\qquad
k_{y}=(\epsilon-v)\sin\phi_{2}+\frac{d_{1}}{l_{B}^{2}}.
\end{equation}
Similarly for the barrier region  $d_{1} \leq x \leq d_{2}$ (region 4) we have
\begin{eqnarray}
&&\epsilon-v=s_{4}
\sqrt{(k_{2x})^{2}+\left(k_{y}+\frac{d_{1}}{l_{B}^{2}}\right)^{2}}\\
%\end{equation}
%\begin{equation}
&&\psi_{4}=\frac{e}{\sqrt{2}}\left(
\begin{array}{c}
1 \\
 z_{4}\end{array}\right)e^{i(k_{4x} x +k_{y} y)}+\frac{f}{\sqrt{2}}\left(
\begin{array}{c}
1 \\
 -z_{4}^*\end{array}\right)e^{i(-k_{4x} x +k_{y}
 y)}
\end{eqnarray}
where $e$ and $f$ are the corresponding wavefunction parameters. Similarly, the momentum can be defined by
\begin{equation}
k_{4x}=(\epsilon-v)\cos\phi_{4},\qquad
k_{y}=(\epsilon-v)\sin\phi_{4}-\frac{d_{1}}{l_{B}^{2}}.
\end{equation}
Finally, in the transmission region, $x > d_{2}$ (region 5), we have only transmitted waves
\begin{equation}
\epsilon=\sqrt{(k_{5x})^{2}+\left(k_{y}+\frac{d_{1}}{l_{B}^{2}}\right)^{2}}
\end{equation}
\begin{equation}
\psi_{5}=\frac{t}{\sqrt{2}}\left(
\begin{array}{c}
1 \\
 z_{5}\end{array}\right)e^{i(k_{5x} x +k_{y} y)}
\end{equation}
where $t$ is the transmission amplitude, the corresponding momentum reads as
\begin{equation}
k_{5x}=\epsilon\cos\phi_{5},\qquad
k_{y}=\epsilon\sin\phi_{5}-\frac{d_{1}}{l_{B}^{2}}.
\end{equation}
The refraction angles $\phi_{2}$, $\phi_{4}$ and $\phi_{5}$ at the interfaces
are obtained by requiring conservation of the momentum $p_{y}$. This leads
to a simplified expression of these angles in terms of $\phi_1$ %as follows
\begin{eqnarray}
\sin\phi_{2} &=&\frac{\epsilon }{\epsilon -v}\sin\phi_{1}\\
\sin\phi_{4} &=& \frac{\epsilon }{\epsilon -v}\sin\phi_{1}+\frac{2d_{1}}{(\epsilon -v)l_{B}^{2}}\\
\sin\phi_{5} &=& \sin\phi_{1}+\frac{2d_{1}}{\epsilon l_{B}^{2}}.
\end{eqnarray}
We should point out at this stage that we were unfortunately forced to adopt a somehow
cumbersome notation for our wavefunction parameters in different potential regions due
to the relatively large number of necessary subscripts and superscripts.

Before matching the eigenspinors at the boundaries,
let us define the following shorthand notation for the dimensionless parameters
\beq
 a_{n \pm}=\sqrt{1\pm
\frac{\mu}{\epsilon_{n}} }, \qquad
d_{n \pm}=\frac{\sqrt{2}}{\epsilon_{n} l_B a_{n \pm} }
\eeq
and the parabolic cylindric functions by
\begin{eqnarray}
\eta_{1 \pm} &=& D_{\left[(\epsilon_{n}l_{B})^{2}-(\mu l_{B})^{2} \right]/2-1} \left[\pm\sqrt{2}
\left(\frac{-d_{1}}{l_{B}}+k_{y}l_{B}\right)\right]\lb{47} \\
\xi_{1 \pm} &=& D_{\left[(\epsilon_{n}l_{B})^{2}-(\mu l_{B})^{2}\right]/2}
\left[\pm\sqrt{2} \left(\frac{-d_{1}}{l_{B}}+k_{y}l_{B}\right)\right].\lb{477}
\end{eqnarray}
Now, requiring the continuity of the spinor wavefunctions at each junction interface give rise to a set of equations. Indeed,
%the related symbols $\eta_{2}^{\pm}$ ,$\xi_{2}^{\pm}$
%follow by letting $-d_{1}\longrightarrow d_{1}$
at the point %$x=-d_{2}$: %we obtain
\begin{eqnarray}\lb{48}
x &=& -d_{2}\ \lga \ \left\{\begin{array}{ll}
  e^{-i k_{1x} d_{2}}+r e^{i k_{1x} d_{2}} =
 a e^{-i k_{2x} d_{2}}+b e^{i k_{2x} d_{2}}\\
   z_{1}e^{-i k_{1x} d_{2}}-rz_{1}^* e^{i k_{1x} d_{2}} =
  az_{2} e^{-i k_{2x} d_{2}}-b z_{2}^*e^{i k_{2x} d_{2}}  \\
\end{array}\right.\\
%\end{equation}
%$x=-d_{1}$:
%\begin{equation}
x &=&-d_{1}\ \lga \ \left\{\begin{array}{ll}
 a e^{-i k_{2x} d_{1}}+b e^{i k_{2x} d_{1}}=
 c_{+}a_{n+}\eta_{1}^{+} +c_{-}a_{n-}\eta_{1}^{-}\\
   az_{2} e^{-i k_{2x} d_{1}}-b z_{2}^*e^{i
   k_{2x} d_{1}}= c_{+}id_{n+}\xi_{1+} -c_{-}id_{n-} \xi_{1-}  \\
\end{array}\right.\\
%\end{equation}
%$x=d_{1}$:
%\begin{equation}
x &=& d_{1}\ \lga \ \left\{\begin{array}{ll}
 c_{+}a_{n+}\eta_{2+} +c_{-}a_{n-}\eta_{2-}= e e^{i k_{4x} d_{1}}
 +f e^{-i k_{4x} d_{1}}\\
c_{+}id_{n+}\xi_{2+} -c_{-}i d_{n-}\xi_{2-}  = e
z_{4}e^{i k_{4x}
d_{1}}-fz_{4}^* e^{-i k_{4x} d_{1}}  \\
\end{array}\right.\\
%\end{equation}
%$x=d_{2}$:
%\begin{equation}\lb{51}
x &=& d_{2}\ \lga \ \left\{\begin{array}{ll}
 e e^{i k_{4x} d_{2}}+f e^{-i k_{4x} d_{2}}=
 t e^{i k_{5x} d_{2}}\\
   ez_{4} e^{ik_{4x} d_{2}}-fz_{4}^*
    e^{-i k_{4x} d_{2}}= tz_{5} e^{i k_{5x} d_{2}} \\
\end{array}\right. \lb{51}
\end{eqnarray}
In the above formulae, the parameters $\eta_{2\pm}$ and  $\xi_{2\pm}$ are defined similarly to (\ref{47})-(\ref{477}) but with $d_1$ replaced by $-d_1$.
All these equations can be written in compact form by introducing
the transfer matrix $M$, which  is given by% ca be written,
\begin{equation}
%M=
\left(%
\begin{array}{llllllllllllllll}
  -e^{ik_{1x}d_{2}} & e^{-ik_{2x}d_{2}} & e^{ik_{2x}d_{2}} & 0 & 0& 0 & 0 & 0 \\
  \frac{1}{z_{1}}e^{ik_{1x}d_{2}} & 1 & \frac{-1}{z_{2}}e^{ik_{2x}d_{2}} & 0 & 0 & 0 & 0 & 0 \\
  0 &  e^{-ik_{2x}d_{1}}& e^{ik_{2x}d_{1}} & -a_{n}\eta_{1+} & -a_{n-}\eta_{1-} & 0 & 0 & 0 \\
  0 &  z_{2}e^{-ik_{2x}d_{1}} & \frac{-1}{z_{2}}e^{ik_{2x}d_{1}} & -id_{n+}\xi_{1+}&
  id_{n-}\xi_{1-} & 0 & 0 & 0 \\
  0 & 0 & 0 & -a_{n+}\eta_{2+} & -a_{n-}\eta_{2-} & e^{ik_{4x}d_{1}} & e^{-ik_{4x}d_{1}} & 0 \\
  0 & 0 & 0 & id_{n+}\xi_{2+} & -id_{n-}\xi_{2-} &-z_{4} e^{ip_{4x}d_{1}} &
  \frac{1}{z_{4}}e^{-ik_{4x}d_{1}} & 0 \\
  0 & 0 & 0 & 0 & 0 & e^{ik_{4x}d_{2}} & e^{-ik_{4x}d_{2}} & -e^{ik_{5x}d_{2}} \\
  0 & 0 & 0 & 0 & 0 &  z_{4}e^{ik_{4x}d_{2}} & -\frac{1}{z_{4}}e^{-ik_{4x}d_{2}} & -z_{5}e^{ik_{5x}d_{2}}  \\
\end{array}%
\right).
\end{equation}
Defining  $\Phi=(r, a, b, c_{+}, c_{-}, e, f, t)^{t}$ as a vector quantity made of all unknown wavefunction parameters in our problem and
$\Xi=(e^{-ik_{1x}d_{2}}, z_{1}e^{-ik_{1x}d_{2}}, 0, 0, 0, 0, 0,
0)^{t}$, where the superscript $t$ stands for transpose, the compact form of equation (\ref{48})-(\ref{51}) can then be written as
\begin{equation}\lb{comf}
M\Phi=\Xi.
\end{equation}
Of course the same problem could be formulated in terms of $2\times 2$ matrices if we use the concept
of transfer matrix from one potential region to another. This last formulation will be much more adequate
in dealing with periodic systems and applying the Bloch theorem to find the associated energy bands.
%This will inspected carefully in the forthcoming analy

%%%%%%%%%%%%%%%%%%%%%%%%%%%%%%%%%%%%%%%%%%%%%%%%%%%%%%%%%%%%%%
\section{Reflection  and transmission coefficients}
%%%%%%%%%%%%%%%%%%%%%%%%%%%%%%%%%%%%%%%%%%%%%%%%%%%%%%%%%%%%%

Now we are ready for the computation of the reflection $R$ and  transmission $T$ coefficients.
For this purpose, we introduce the associated current density $J$, which defines $R$ and $T$ as
\beq\lb{RTT}
R=\frac{J_{\sf re}}{ J_{\sf in}}, \qquad T=\frac{ J_{\sf tr}}{ J_{\sf in}}
\eeq
%Using the reflected $J_{refl}$ and transmitted $J_{trans}$
%currents,
 %we have that the reflection and transmission coefficients
%R and T can be expressed as
%\begin{equation}
 % T=\frac{ J_{trans}}{ J_{inc}},\qquad R=\frac{J_{refl}}{ J_{inc}}
%\end{equation}
%The transmission probability T and reflection
%probability R are not independent, the are related via the
%unitarity condition T+R=1.\\
where $J_{\sf in}$, $J_{\sf re}$ and $J_{\sf tr}$ stand for the incident,
reflected and transmitted components of the current density, respectively.
It is easy to show that the current density $J$ is given by
%The electrical current density J is given by the expression
\begin{equation}
J= ev_{\sf F}\psi^{\dagger}\sigma _{x}\psi
\end{equation}
which gives the following results for the incident, reflected and transmitted components
\begin{eqnarray}
J_{\sf in} &=&  ev_{\sf F}(\psi_{1in})^{\dagger}\sigma_{x}\psi_{1in}=ev _{\sf F}\frac{k_{1x}}{\sqrt{(k_{1x})^{2}+
  \left(k_{y}-\frac{d_{1}}{ l_{B}^{2}}\right)^{2}}}\\
  J_{\sf re} &=& e v_{\sf F} (\psi_{1re})^{\dagger}\sigma _{x}\psi_{1re}=
  ev_{\sf F}r^{*}r\frac{k_{1x}}{\sqrt{(k_{1x})^{2}+
  \left(k_{y}-\frac{d_{1}}{ l_{B}^{2}}\right)^{2}}}\\
 J_{\sf tr} &=& e v_{\sf F}(\psi_{5tr})^{\dagger}\sigma _{x}\psi_{5tr}
 =e v_{\sf F}t^{*}t\frac{p_{5x}}{\sqrt{(k_{5x})^{2}+
  \left(k_{y}+\frac{d_{1}}{l_{B}^{2}}\right)^{2}}}.
\end{eqnarray}

Injecting these results in (\ref{RTT}) we obtain
\begin{eqnarray}
  T &=& \frac{k_{5x}}{k_{1x}}\frac{\sqrt{(k_{1x})^{2}+\left(k_{y}-
  \frac{d_{1}}{l_{B}^{2}}\right)^{2}}}{\sqrt{(k_{5x})^{2}+
  \left(k_{y}+\frac{d_{1}}{l_{B}^{2}}\right)^{2}}} |t|^{2}\\
  R &=& |r|^{2}.
\end{eqnarray}
Now using the conservation of energy
\begin{equation}
\sqrt{(k_{1x})^{2}+\left(k_{y}-\frac{d_{1}}{l_{B}^{2}}\right)^{2}}=
\sqrt{(k_{5x})^{2}+\left(k_{y}+\frac{d_{1}}{l_{B}^{2}}\right)^{2}}
\end{equation}
we find the constraint
\begin{equation}
k_{1x}=\sqrt{(k_{5x})^{2}+4k_{y}\frac{d_{1}}{ l_{B}^{2}}}
\end{equation}
which allows us to express the transmission probability $T$ in the following form
\begin{equation}
  T= %\frac{p_{x5}}{p_{x1}}\mid t\mid^{2}=
  \frac{k_{5x}}{\sqrt{(k_{5x})^{2}+4k_{y}\frac{d_{1}}{ l_{B}^{2}}}} |t|^{2}.
\end{equation}
To get an explicit expression for $T$, we should determine the transmission amplitude $t$.
After some lengthy algebra, one can solve the linear system of
equations (\ref{comf}) to obtain the transmission and reflection amplitudes in closed form
\begin{eqnarray}
  t &=& \frac{i\sqrt{2}}{l_B|\epsilon-u|}\frac{e^{i(d_{1}+d_{2})(k_{2x}+k_{4x})}
  (1+(z_{1})^{2})(1+(z_{2})^{2})(1+(z_{4})^{2})}
  {e^{id_{2}(k_{1x}
  +k_{5x})} \left[A(1+z_{1}z_{2})+B(z_{1}-z_{2})\right]} \left(\eta_{2+}\xi_{2-}+\eta_{2-}\xi_{2+}\right)\\
  r &=& \frac{e^{-2id_{2}k_{1x}}z_{1}\left[  A(z_{1}-z_{2})-B(1+z_{1}z_{2})\right]}
  {A(1+z_{1}z_{2})+B(z_{1}-z_{2})}
\end{eqnarray}
where we have defined the following quantities
\begin{eqnarray}
  A &=& e^{2i(d_{1}k_{2x}+d_{2}k_{4x})}(z_{4}-z_{5})(\alpha+\beta
  z_{2} z_{4}+\gamma z_{4}-\delta z_{2})\\
  && +e^{2id_{1}(k_{2x}+p_{4x})}(1+z_{4}z_{5}) \left(-\alpha
  z_{4}+\beta z_{2}+\gamma+\delta z_{2} z_{4}\right) \nonumber\\
  B &=& e^{2id_{2}(k_{2x}+k_{4x})}(z_{4}-z_{5})\left(-\alpha
  z_{2}+\beta z_{4}-\gamma z_{2}z_{4}-\delta \right)\\
  &&+e^{2i(d_{2}k_{2x}+
  d_{1}k_{4x})}(1+z_{4}z_{5}) \left(\alpha z_{2}z_{4}+\beta-\gamma
  z_{2}+\delta z_{4}\right)\nonumber
  \end{eqnarray}
as well as
  \begin{eqnarray}
\alpha &=& \frac{\epsilon -u+\mu }{\epsilon
-u} \left(\eta_{1-}\eta_{2+}-\eta_{1+}\eta_{2-}\right)\\
 \beta &=& \frac{2}{l_B(\epsilon
-u)(\epsilon -u+\mu )}
\left(\xi_{1-}\xi_{2+}-\xi_{1+}\xi_{2-} \right)\\
 \gamma &=& i\frac{\sqrt{2}}
  {l_B|\epsilon
-u |} \left(\eta_{1-}\xi_{2+}+\eta_{1+}\xi_{2-} \right)\\
\delta &=& i\frac{\sqrt{2}}
  {l_B|\epsilon -u |} \left(\eta_{2-}\xi_{1+}+\eta_{2+}\xi_{1-}\right).
\end{eqnarray}
The resulting reflection and transmission coefficients have been computed numerically and
are plotted in Figures 3, 4, 5, 6 for several parameter values ($\epsilon$, $v$, $u$, $d_{1}$, $d_{2}$, $\mu$). For instance
a typical value of the magnetic field say $B_{0}=4 {\sf T}$, the magnetic length is $l_{B} = 13$ {\sf nm}, and $\epsilon l_{B}=1$ corresponds
to the energy $E = 44 ~ {\sf meV}$ \cite{demartino}.

Now, let us study the reflection and transmission coefficients versus the energy $\epsilon l_{B}$.
The quantity $k_{y} l_{B}= m^{*}$ plays a very important role in the
transmission of Dirac fermions via the obstacles created by the
series of scattering potentials, because it associates an effective mass to the particle and hence determines the threshold for the
allowed energies. However,  the application of the magnetic field in the intermediate zone seems to reduce this effective mass to
$\left(k_{y}l_{B}-\frac{d_{1}}{l_{B}}\right)$ in the incidence region while it increases it to
$\left(k_{y}l_{B}+\frac{d_{1}}{l_{B}}\right)$ in the transmission region as shown in Figure 2. The allowed energies
are then determined by the greater effective mass, namely
$\epsilon l_{B}\geq k_{y}l_{B}+\frac{d_{1}}{l_{B}}$.\\

\begin{center}
  \includegraphics[width=5in]{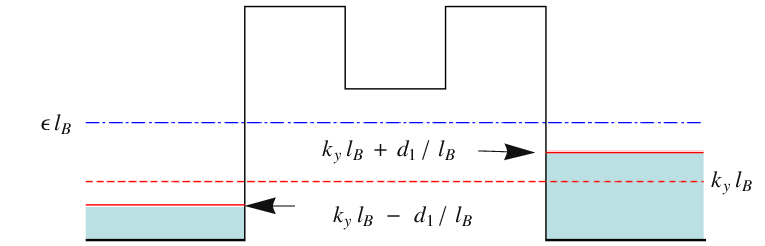}\\
\end{center}
{{\sf Figure 2: The energy configuration of the double barrier potential
illustrating the effect of inhomogeneous effective mass in the incidence and reflection regions.  "Color figure online".}}\\
%\end{center}
%\end{figure}

If we cancel the well region by setting $d_1=0$ then we reproduce the usual single barrier transmission as reflected in figure 3
which shows clearly the Klein zone followed by a full reflection zone. On the other hand if we keep the well region and cancel
both the applied magnetic field and mass term in the well region, the series of potentials
behaves like a simple double barrier with the same effective mass $k_{y}$ all over.
Thus, in this case we reproduce exactly the transmission obtained in~\cite{abj}, for the massive 1D Dirac
equation with $m = k_{y}$.\\

%(see fig )traced by
%our calculation that attaches  exactly with that found in
%references [....]
\begin{center}
  \includegraphics[width=4in]{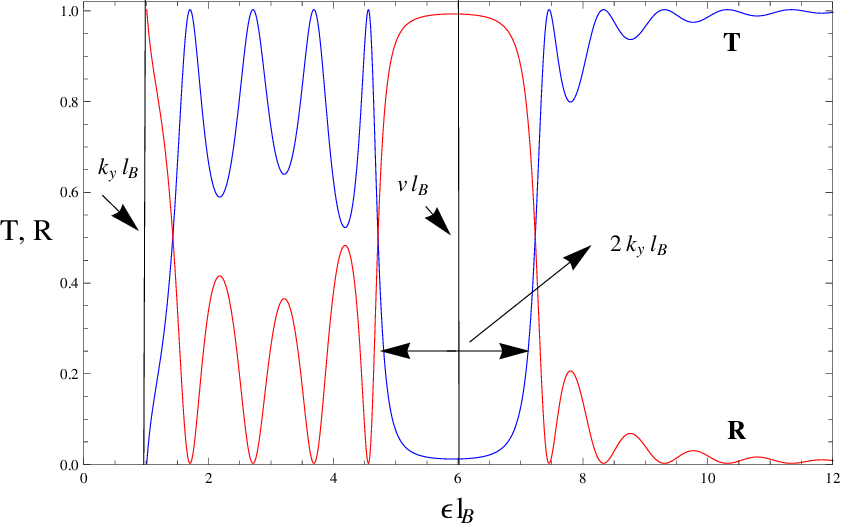}
\end{center}

%\begin{center}
 {\sf {Figure 3: Transmission and reflection coefficient as a function of energy
 $\epsilon l_{B}$ for a single barrier with $\frac{d_{1}}{l_{B}}= 0 $, $d_{2} l_{B}= 1.5$, $v l_{B}= 6$ and $k_{y} l_{B}= 1$.  "Color figure online".}}\\
%\end{center}
%\end{figure}

\noindent The result shown in Figure 3 can be interpreted as follows. The transmission
as function of energy, through a single barrier, starts from the minimum required energy equal
to the effective mass $ m^{*}=k_yl_B$ then oscillates
reaching full transmission due to constructive interference in the (Klein zone) for energies between
$ m^{*}$ and $V-m^{*}$. This is then followed by a gap region between $V-m^{*}$ and $V+m^{*}$ where the transmission
drops to zero. For energies above $V+m^{*}$ the transmission has the usual oscillating behavior then reaches full
total transmission asymptotically.

In Figure 4 we show the influence of the size of the well region and magnetic field on the transmission.
First we consider the same conditions as above but we vary $d_{1}$ to obtain Figure 4a:\\
\begin{center}
\includegraphics[width=4in]{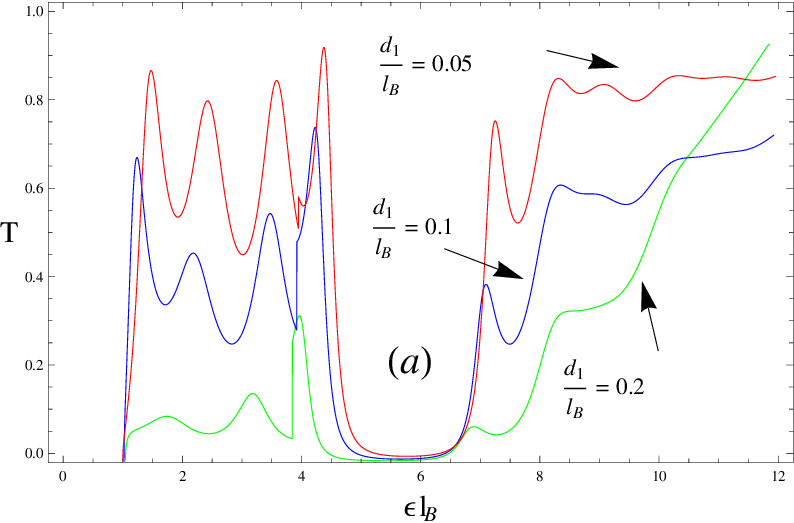}
\end{center}

%\begin{center}
 {\sf{Figure 4a: Transmission  as a function of energy $\epsilon l_{B}$ with
   $\frac{d_{1}}{l_{B}}=\{0.05,\ 0.1,\ 0.2\}$,  $\frac{d_{2}}{l_{B}}=1.5$, $V l_{B}=6$, $u l_{B}=4$, $\mu l_{B}=4$, $k_{y}l_{B}=1$.  "Color figure online".}}\\

 \noindent According to Figure 4, the size of the intermediate zone affects the form of transmission
in these four zones. As $d_{1}$ increases, the effective mass increases and the oscillations in the Klein
zone get reduced. This strong reduction in the transmission in the Klein zone seem to suggest the potential
suppression of the Klein tunneling as we increase $d_1$.
On the other hand the size of the gap region increases as we increase $d_1$. However, this is not
the case as we increase $\mu l_{B}$ where there is some change amplitude as shown in  Figure 4b:\\
\begin{center}
\includegraphics[width=4in]{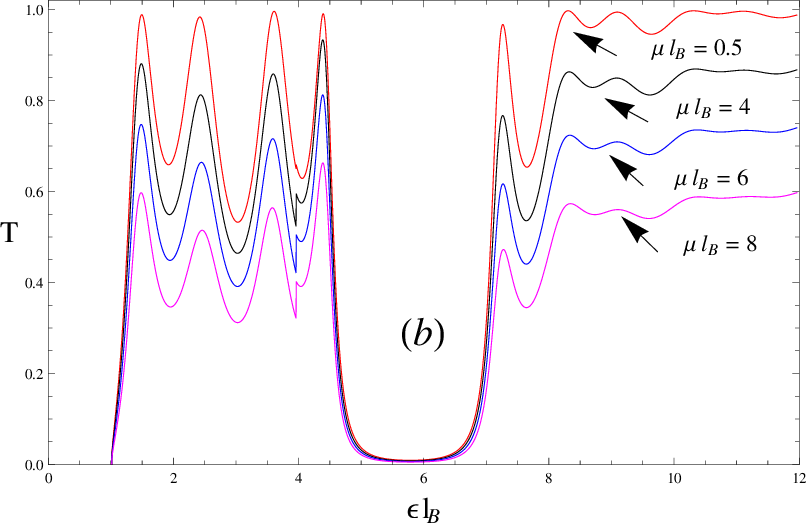}
\end{center}

%\begin{center}
 {\sf{Figure 4b: Transmission  as a function of energy $\epsilon l_{B}$ with
   $\mu l_{B}=\{0.5,\ 4,\ 6,\ 8\}$,   $\frac{d_{1}}{l_{B}}=0.05$, $\frac{d_{2}}{l_{B}}=1.5$, $v l_{B}=6$, $u l_{B}=4$, $k_{y}l_{B}=1$.  "Color figure online".}}\\

%%%%%%%%%%%%%%%%%%%%%%%%%%%%%%%%%%%%%%%%%%%%%%%%%%%
%\subsubsection{ Transmission T and reflection coefficient R versus the potential $V l_{B}$}
%%%%%%%%%%%%%%%%%%%%%%%%%%%%%%%%%%%%%%%%%%%%%%V

It is worthwhile to analyze the reflection and transmission coefficients versus the potential.
In doing so, we fix the energy and choose a value of $d_{1}$, then we compute the transmission as shown in Figure 5a
and 5b:\\
\begin{center}
\includegraphics[width=4in]{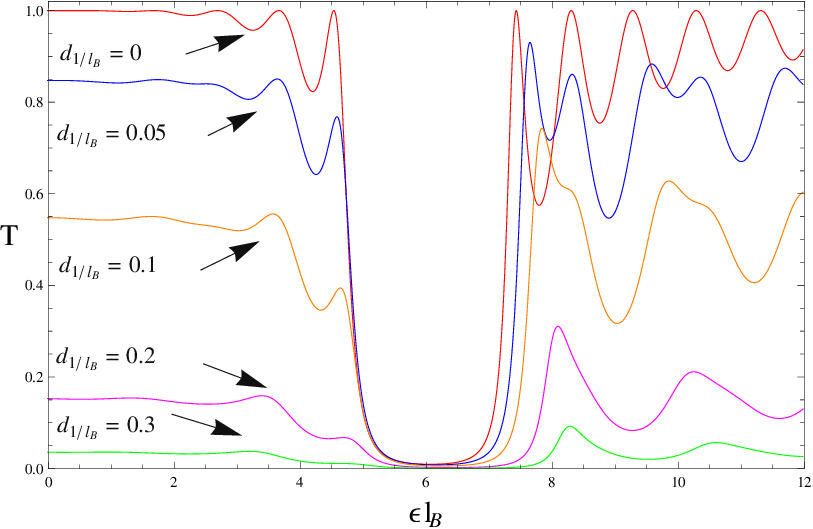}
\end{center}

%\begin{center}
 {\sf{Figure 5a: Transmission  as a function of potential  $vl_{B}$ for
   $\frac{d_{1}}{l_{B}}=\{0,\ 0.05,\ 0.1,\ 0.2,\ 0.3\}$  with $\frac{d_{2}}{l_{B}}=1.5$, $\epsilon l_{B}=6$, $ul_{B}=4$, $\mu l_{B}=4$ and $k_{y}l_{B}=1$.  "Color figure online".}}
\begin{center}
\includegraphics[width=4in]{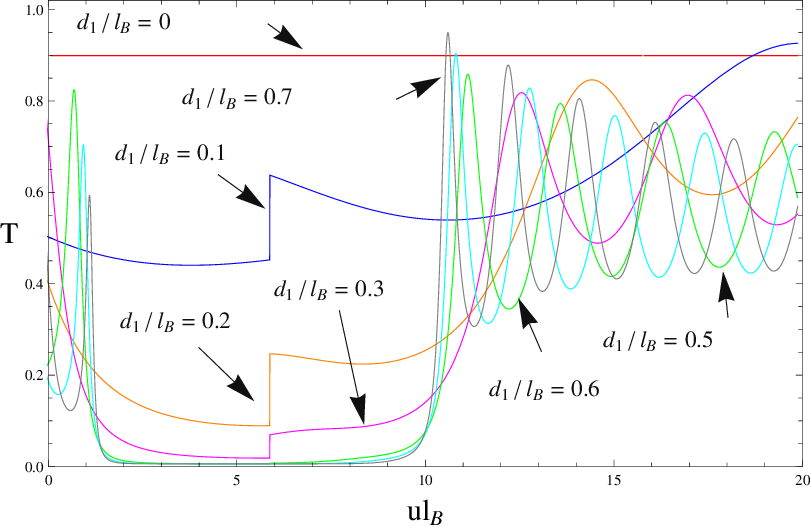}
\end{center}

%\begin{center}
 {\sf{Figure 5b: Transmission  as a function of potential  $ul_{B}$ for
   $\frac{d_{1}}{l_{B}}=\{0,\ 0.1,\ 0.2,\ 0.3,\ 0.5,\ 0.6, \ 0.7\}$
   with $\frac{d_{2}}{l_{B}}=1.5$, $\epsilon l_{B}=6$, $vl_{B}=4$, $\mu l_{B}=4$ and $k_{y}l_{B}=1$. "Color figure online".}}\\

\noindent It is clearly seen that for a given energy, the transmission decreases if $d_{1}$ increases and then vanishes.

Let us examine the transmission coefficient as function of angle incidence  $\phi_{1}$. This
is shown in Figure 6:\\
\begin{center}
\includegraphics[width=1.5in]{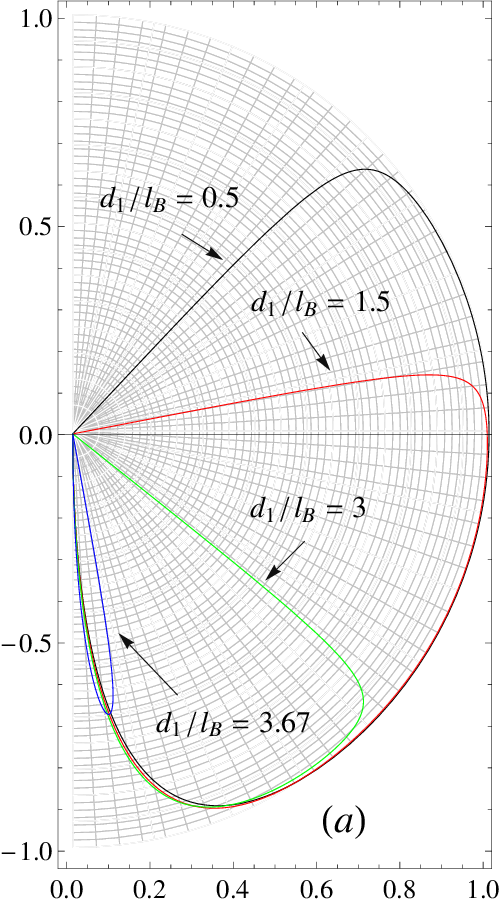}\ \ \ \ \ \ \ \ \ \ \ \ \ \ \ \
\includegraphics[width=1.5in]{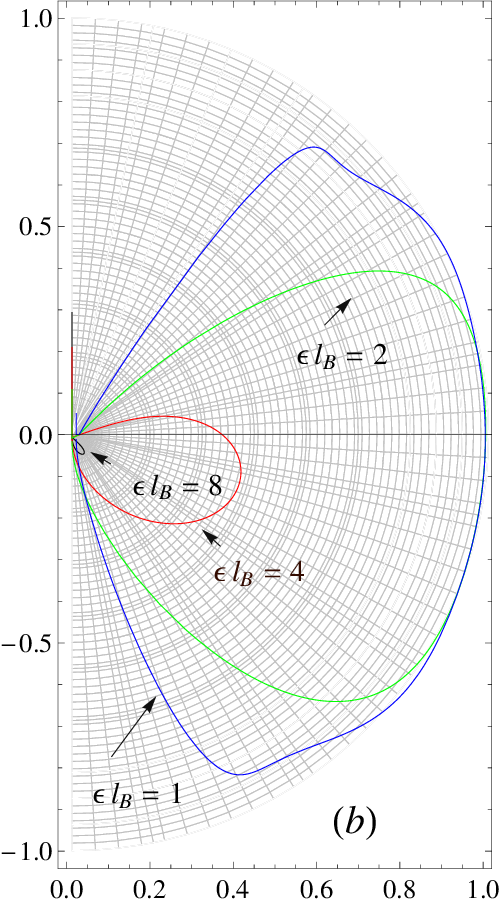}
\end{center}

{\sf{Figure 6a,6b: Polar plot of a curve with radius (transmission T) as a function of angle $\phi_{1}$ with:\\
 (a): $\frac{d_{1}}{l_{B}}=\{0.5,\ 1.5,\ 3, \ 3.67\}$ with $\epsilon l_{B}=3.7$, $\frac{d_{2}}{l_{B}}=\frac{d_{1}}{l_{B}}$,
 $v l_{B}=u l_{B}=0$, $\mu l_{B}=0$, $k_{y}l_{B}=1$.}}\\
 {{\sf (b): $\epsilon l_{B}=\{1,\ 2,\ 4, \ 8\}$ with $\frac{d_{1}}{l_{B}}=0.5$, $\frac{d_{2}}{l_{B}}=0.6$, $v l_{B}=0.5$, $u l_{B}=0.4$, $\mu l_{B}=1$,
  $k_{y}l_{B}=1$. \\ "Color figure online".}}\\

\noindent Figure 6a reproduces exactly the result of De Martino \emph{et al}.~\cite{demartino},
this reference was the first to treat the confinement of Dirac fermions by an inhomogeneous
magnetic field. This polar graph shows the transmission as a function of the incidence angle,
the outermost circle corresponds to full transmission, $T = 1$, while the origin of this plot
represents zero transmission. We note that for certain incidence angles the transmission is not allowed,
in fact for $\epsilon l_B \leq k_y l_B + \frac{d_1}{l_B}$ all waves are completely reflected.
It is worth mentioning that the transmission is uniquely defined by the incidence angle i.e.
each radial line representing a given incidence angle intersects the transmission curve at one point.
In Figure 6a we see that the transmission vanishes for values of $\frac{d_1}{l_B}$ below the critical value $\epsilon l_B = 6.7$.
 In Figure 6b we fix $\frac{d_1}{l_B}$ and $\frac{d_2}{l_B}$ and vary the energy, we observe that the transmission get reduced as
 we increase the increase the energy. In Figure 6c we fix  $\frac{d_2}{l_B}$ and vary $\frac{d_1}{l_B}$,
 the transmission vanishes below the critical value $\frac{d_1}{l_B} = 3.7$ for our parameters. \\
\begin{center}
  \includegraphics[width=1.5in]{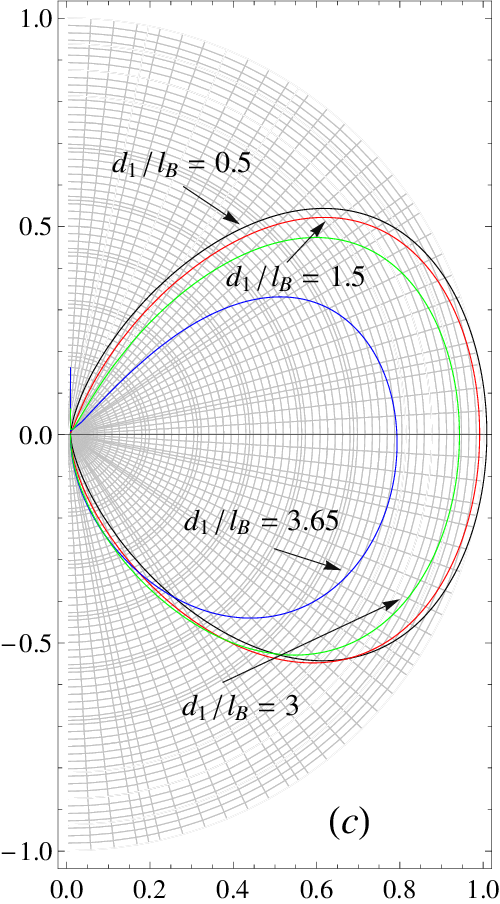} \ \ \ \ \ \ \ \ \ \ \ \ \ \ \ \
  \includegraphics[width=1.5in]{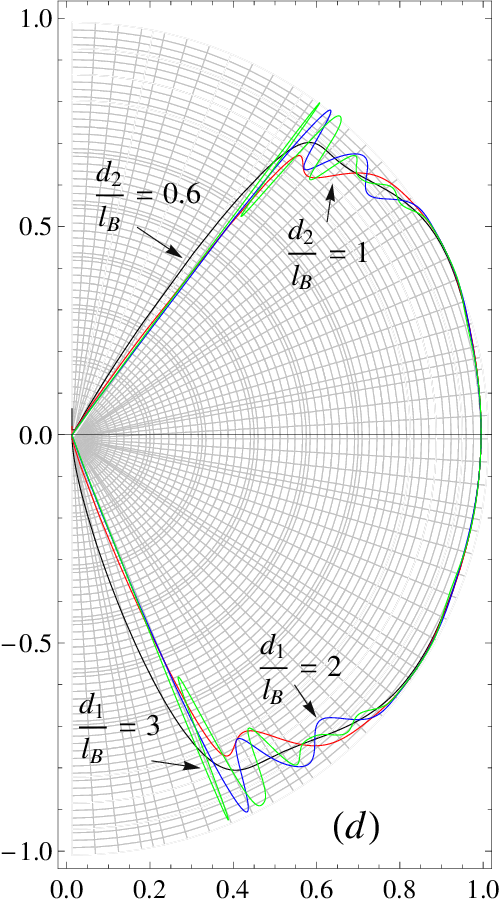}
\end{center}
  {\sf{Figure 6c,6d: Polar plot of a curve with radius (transmission T) as a function of angle $\phi_{1}$ with:}}\\
{\sf{
 (c): $\frac{d_{1}}{l_{B}}=\{0.5,\ 1.5,\ 3, \ 3.67\}$ with $\epsilon l_{B}=2$, $\frac{d_{2}}{l_{B}}=0.6$, $v l_{B}=0.5$, $u l_{B}=0.4$, $\mu l_{B}=1$,
 $k_{y}l_{B}=1$.}}\\
{{\sf (d): $\frac{d_{2}}{l_{B}}=\{0.6,\ 0.1,\ 2, \ 3\}$ with $\epsilon l_{B}=8$, $\frac{d_{2}}{l_{B}}=0.5$,
$v l_{B}=0.5$, $u l_{B}=0.4$, $\mu l_{B}=1$, $K_{y}l_{B}=1$. \\ "Color figure online".}}

\begin{center}
  \includegraphics[width=1.5in]{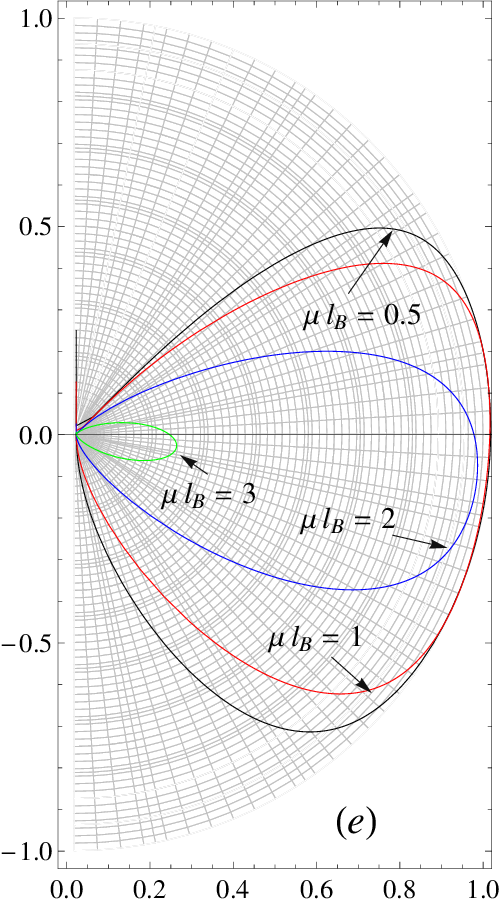}
\end{center}
{\sf{Figure 6e: Polar plot of a curve with radius (transmission T) as a function of angle $\phi_{1}$ with :}}

{\sf {(e): $\mu l_{B}=\{0.5,\ 1,\ 2, \ 3\}$ with $\epsilon l_{B}=4$, $\frac{d_{1}}{l_{B}}=0.5$,
$\frac{d_{2}}{l_{B}}=0.6$, $v l_{B}=0.5$, $u l_{B}=0.4$, $K_{y}l_{B}=1$.  "Color figure online".}}\\

 \noindent In Figure 6d we fix  $\frac{d_1}{l_B}$
 and vary $\frac{d_2}{l_B}$, the transmission is less sensitive to the variations of $\frac{d_2}{l_B}$
 and shows an oscillatory behavior which becomes more apparent for larger values of $\frac{d_2}{l_B}$.
 Finally in Figure 6e we show how the transmission is affected by the effective mass term reflected by $\mu$,
 the transmission decreases sharply as we increase $\mu$.

%%%%%%%%%%%%%%%%%%%%%%%%%%%%%%%%%%%%%%%%%%%%%%%%%%%%%%
\section{Conclusion}
%%%%%%%%%%%%%%%%%%%%%%%%%%%%%%%%%%%%%%%%%%%%%%%%%%%%%%%%%

To conclude, we have studied the effect of gap opening on the transport properties of graphene Dirac fermions.
The effective mass is generated by a lattice miss-match between the boron nitride substrate and the 2D graphene sheet.
An additional homogeneous magnetic field localized within the well potential region causes the effective mass
to change and affects asymmetrically the incidence and reflection regions.

We have found that in contrast to electrostatic barriers, magnetic barriers are able confine Dirac fermions, a property
of great importance from the practical point of view. In fact as can be seen in Figure 6, the transmission reduces a lot
as we increase the effective mass parameter, $t'$ in our original model (\ref{31}), or the width, $d_1$, of the effective mass region.
These results were exposed clearly on the polar graph showing the transmission as a function of the incidence angle in Figure 6.
The outermost semicircle corresponds to unit transmission is not intersected by the transmission curve, even at normal incidence,
if we increase the effective mass or width of the magnetic field region.

We hope that the present interplay between gap opening and local magnetic field effect will give more
freedom to experimentalists to develop graphene based electronic devices.

%%%%%%%%%%%%%%%%%%%%%%%%%%%%%%%%%%%%%%%
\section*{Acknowledgments}
%%%%%%%%%%%%%%%%%%%%%%%%%%%%%%%%%%%%%%%%

The generous support provided by the Saudi Center for Theoretical Physics (SCTP)
is highly appreciated by all Authors. AJ acknowledge partial support by King Faisal
University. We also acknowledge the support of KFUPM
under project RG1108-1-2.

\end{document}